# Towards optimization of pulsed sodium laser guide stars

Rachel Rampy,[1,*] Donald Gavel,[1] Simon M. Rochester,[2] Ronald Holzlöhner[3]

[1]University of California, 1156 High St., Santa Cruz, CA, 95064, USA
[2]Rochester Scientific, LLC, El Cerrito, CA 94530, USA
[3]European Southern Observatory (ESO), Garching bei München, D-85748, Germany
*Corresponding author: rrampy@keck.hawaii.edu

**Pulsed sodium laser guide stars (LGS) are useful because they allow for Rayleigh blanking and fratricide avoidance in multiple-LGS systems. Bloch-equation simulations of sodium-light interactions show that these may be able to achieve photon returns nearly equal to, and in some cases greater than, what is seen from continuous-wave (CW) excitation. In this work, we study the time-dependent characteristics of sodium fluorescence, and investigate the optimal format for the new fiber laser LGS that will be part of the upgraded adaptive optics (AO) system on the Shane telescope at Mt. Hamilton. Results of this analysis are examined in the context of their general applicability to other LGS systems and the potential benefits of uplink correction are considered. Comparisons of simulation predictions with measurements from existing LGS are also presented and discussed.**

*OCIS codes:* (140.0140) Lasers and laser optics, (350.1260) Astronomical optics, (020.0020) Atomic and molecular physics

## 1. INTRODUCTION

Use of laser guide stars (LGS) along with adaptive optics (AO) allows large telescopes to significantly improve resolution by correcting image distortion induced by atmospheric turbulence at nearly any location in the sky. Sodium LGS operating at a wavelength of 589 nm employ resonance fluorescence of mesospheric sodium atoms, which are concentrated in a ~10 km-thick layer, centered at an altitude of ~90 km. These are preferable to Rayleigh guide stars due to their high altitude (allowing a larger fraction of the turbulence to be measured). Sodium is preferred as a resonant absorber, compared to the other constituents of the upper atmosphere, since it fluoresces at visible wavelengths and has a large fluorescence cross section–abundance product (Happer et al. 1994). Because powerful diffraction-limited laser beams at 589 nm are expensive to produce, careful attention must be paid to optimizing the laser format (spectrum, polarization, spot size, pulse length and duty cycle) in order to maximize the return signal.

Numerical models allow investigation of the various physical mechanisms occurring in the laser–sodium system that affect the number of photons returned to the telescope. The investigations presented here use a density-matrix calculation with coupled velocity groups, and include the important physical effects of Larmor precession due to the geomagnetic field, radiation pressure (recoil), saturation, and velocity-changing and spin-randomizing collisions (Holzlöhner et al. 2010a). The LGSBloch package runs in Mathematica, and is based on the Atomic Density Matrix package written by S. Rochester*.

Motivation for this research was provided by the upgrade of the AO system on the 3 meter Shane Telescope at Lick Observatory. The Shane hosted the first experiments in sodium LGS AO, with observations starting in 1996, and provides for regular astronomical science observing to this day. The upgraded system incorporates many of the recent advancements in AO technology and lessons learned from laboratory and on-sky experiments in order to provide higher Strehl, improved sensitivity, and greater wavelength coverage for astronomers (Kupke et al. 2012). This second-generation system uses a 32×32 actuator micro-electro-mechanical systems (MEMS) deformable mirror, with a higher sensitivity wavefront sensor (WFS). A new fiber laser that was developed at Lawrence Livermore National Laboratory (LLNL) will soon replace the existing dye laser. The fiber technology is more compact and energy efficient than the current laser, and allows adjustments to the pulse and spectral format for optimal coupling to the mesospheric sodium atoms. The initial and principal goal of this investigation has been to ascertain the ideal values of these parameters, and to estimate the return signal this new laser will produce.

The next section presents the sodium physics relevant to the numerical model, followed by description of endeavors to find the optimal pulse and spectral formats in Section 3. Section 4 addresses the potential benefits of uplink correction of the laser beacon, and gives the projected return flux for the new LGS at Lick Observatory. Section 5 presents simulation results of the return flux for the 50 W Fasortronics laser at SOR, the current 10 W dye laser on the Shane telescope, and the two ~15 W LGS at the Keck Observatory. In section 6, an idea that may enable pulsed lasers to create brighter LGS than CW is presented. The final section summarizes what has been determined thus far and briefly discusses where future efforts should focus.

## 2. UNDERSTANDING AND MODELING SODIUM-LIGHT INTERACTIONS

Sodium LGS take advantage of the $3^2S_{1/2}$–$3^2P_{3/2}$ transition in atomic sodium, known as the $D_2$ line. The ground state consists of two hyperfine multiplets, with 8 magnetic substates combined. The hyperfine states are separated by 1.77 GHz, splitting the $D_2$ line into the $D_2a$ and $D_2b$ transition groups. These two groups are composed, respectively, of transitions originating from the $F = 1$ and $F = 2$ ground states, where $F$ is the total atomic angular momentum quantum number. The four excited state multiplets ($F = 0, …, 3$) are separated by 16, 34, and 60 MHz from each other, and contain a total of 16 substates. Figure 1 shows the atomic structure of sodium starting from the Bohr model (on the left) and ending with the hyperfine states and energy separations on the far right (Hillman et al. 2008). At mesospheric temperatures of about 190 K, the motions of the atoms result in the $D_2a$ and $D_2b$ lines being Doppler broadened to ~1 GHz each, giving rise to the characteristic two-peaked absorption/fluorescence profile.

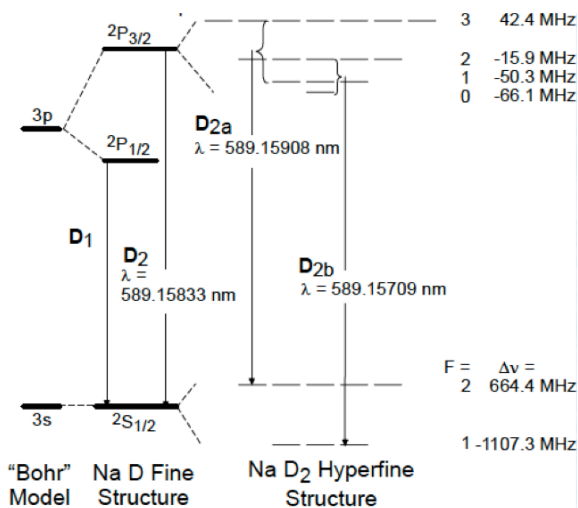

Fig. 1. The atomic structure of sodium, starting from the Bohr model of the atom on the left, and ending on the right with the hyperfine states of the $D_2$ line. The 1.77 GHz separation of the ground states requires that LGS address both transition groups for optimal coupling to the light field (Hillman et al. 2008).

Circularly polarized laser light can induce higher return flux than linearly polarized light due to the phenomenon of optical pumping (Jeys et al. 1992). When an atom encounters a circularly polarized photon resonant with a transition to the excited state, the value of the magnetic quantum number, $m$, must change by $\pm 1$ depending on the handedness of the light, as required by conservation of angular momentum. After several cycles of excitation and spontaneous emission (the excited state lifetime is 16.2 ns), the atoms will tend to be locked cycling between the $F = 2, m = \pm 2$ ground state and the $F' = 3, m = \pm 3$ excited state. This confers three main benefits: this cycling transition has the highest cross section of the entire $D_2$ group, atoms are forbidden by selection rules from decaying from the $F' = 3$ excited state to an $F = 1$ ground state, and the florescence is directed preferentially along the light beam (as for radiation from a dipole rotating about the laser beam axis). However, due to ground-state relaxation and Larmor precession there is never complete polarization in the ground state, resulting in some atoms being excited on other transitions besides the cycling transition. Additionally, spin-exchange collisions tend to equalize the populations between the $F = 1$ and $F = 2$ ground-state hyperfine levels, but this mechanism is slow enough that repumping is still beneficial.

Based on theoretical and experimental considerations, maximizing the number of photons emitted back towards the telescope involves the use of a circularly polarized beam centered on the $D_2a$ line, with ~10% of the light tuned to the $D_2b$ transition (Milonni et al. 1998, Holzlöhner et al. 2010a). With this "repumping," the fraction of atoms that later decay into the $F = 1$ ground state and thus cannot be re-excited by laser light in $D_2a$ is strongly reduced. This optical pumping with repumping strategy allows the most atoms to become trapped in the aforementioned desirable two-state system. Due to the optical pumping effect, the return flux per unit irradiance at first increases with higher laser irradiance. As the irradiance continues to be increased, however, the efficiency eventually falls off, in part due to the increasing fraction of atoms in the excited state (transition saturation). A sodium atom in the excited state that encounters a laser photon may undergo stimulated emission, transitioning back to the ground state by emitting a photon along the laser propagation direction (i.e. off into space). Because of this, the highest efficiency is found near ~10–100 W/m² (for 12% repumping). This peak in efficiency represents the middle ground between having high enough light intensity to trap as many atoms as possible into cycling on the two-state optical pumping transition, without causing a substantial fraction to undergo stimulated emission.

There are three main factors that reduce the coupling efficiency of the laser light to the sodium atoms: (1) Larmor precession due to the geomagnetic field, (2) collisions with other constituents of the upper atmosphere, and (3) the aforementioned transition saturation leading to stimulated emission (Holzlöhner et al. 2010a). Also important to consider is the slight change in velocity incurred by an atom from absorption and spontaneous emission of a photon (equivalent to a Doppler shift of ~50 kHz, known as atomic recoil), and the transit of atoms in and out of the light beam. At the location of Mt. Hamilton, the strength of the geomagnetic field is 0.47 Gauss at 90 km altitude, corresponding to a Larmor precession period of ~3.1 µs. The time between collisions is on the order of ~100 µs, and the atoms in the beam are exchanged on a timescale of several milliseconds. Thus, different inhibitory factors will dominate depending on the duration of the laser pulse.

The quantity of returned photons also depends on the amount of atmospheric absorption and the column density of sodium atoms in the mesosphere. These give rise to a directional dependence factor of $T_a^{2X}/X$, where $T_a$ is the one-way atmospheric transmission at 589 nm at zenith, and $X = 1/\cos(\zeta)$ is the airmass with the zenith angle $\zeta$. For circularly polarized light, the local geomagnetic field also causes significant directional dependence, with maximal reduction of the return flux when the magnetic field and the laser beam are perpendicular. The highest returns are achieved when the laser propagates parallel to the magnetic field. The Na column density is also variable, including a seasonal variation by a factor of ~3 at North American latitudes, with the high around November and low near May (Gumbel et al. 2007).

Modeling sodium-light interactions and predicting LGS brightness also requires accounting for atoms with different Doppler shifts. To accomplish this, the velocity dependence of the density matrix is discretized to describe an appropriate number, $n_{vg}$, of velocity groups each with a fixed component of velocity along the laser beam propagation direction. Because coherences between atoms with different velocities can be neglected, the complete density matrix can be thought of as a collection of $n_{v.g.}$ separate, but coupled, density matrices, each of dimension 24 × 24, since there are a total of 24 magnetic substates in the D2 transition. In the freely available version of LGSBloch, the system is solved using methods built into Mathematica, based on matrix exponentials (described in detail in Holzlöhner et al. 2012). The fluorescent photon flux per solid angle emitted in a given direction is then found from the steady-state solution as the expectation value of a fluorescence operator (Corney 2006, Holzlöhner et al. 2010a).

To account for a non-uniform irradiance pattern in the mesosphere, LGSBloch assumes a Gaussian intensity distribution, discretized with a user-defined number of irradiance levels. Based on computational experimentation, results do not change significantly when greater than five levels are modeled, so that number is used throughout this work. For the most general format comparisons, results are presented with units of specific return, a measure of local return flux per laser irradiance in the mesosphere, as photons/s/sr/atom/(W/m$^2$). When quantities such as the irradiance and/or sodium abundance are held constant the units change to reflect that. The vertical distribution of sodium atoms is modeled as Gaussian, and the code used by Rochester Scientific accounts for variations in collision rates with altitude (although this has been found to have a negligible effect). In calculating the photon flux present at a telescope, the fluorescing atoms are assumed to be at the median altitude of the sodium layer near 92 km.

## 3. OPTIMIZATION OF PULSE LENGTH AND SPECTRAL FORMAT

The first ~30 km of atmosphere strongly scatters light back towards the telescope (Rayleigh scattering), possibly disturbing measurement of the wavefront from the higher-altitude sodium beacon. Rayleigh blanking involves blocking light from entering the WFS during the time the pulse is traveling through this region of air. Fratricide refers to the Rayleigh backscattered light from one LGS entering the WFS of another in systems that contain multiple lasers (such as the Thirty Meter Telescope, Wang et al. 2010), and can be reduced with similar time-gating schemes, or else by launching the laser beams from the side of the primary mirror, as in the European Extremely Large Telescope (E-ELT, Bonaccini-Calia et al. 2014). These background mitigation techniques reduce the WFS measurement error, and thus increase AO system performance.

Until recently, the available laser technology at wavelengths of 589 nm relied heavily on the high peak power in short-pulsed systems to achieve acceptable frequency conversion efficiency levels. As new technologies emerge with greater flexibility available in the temporal format, understanding the costs and benefits of the different format options becomes crucial.

Efforts to determine the optimal pulse duration began with examining how the number of spontaneously emitted photons varies as a function of time. Figure 2 shows results from two 100 μs simulations for a narrow line width laser, zenith pointing at Mt. Hamilton, with the irradiance levels set to be equal to what is present for 10 W average launched power operating at a 20% duty cycle. In the top graph, all the light is on the D$_2$a transition, and the lower plot is for ~10% of the light tuned to D$_2$b. For both cases there is an initial enhancement in the specific return (i.e. photons/s/sr/atom/(W/m$^2$) in the mesosphere). However, in the case of no repumping, the decline begins sooner for all but the lowest irradiance levels, and the peaks are considerably lower. These results suggest that for 10 W average launched power and 20% duty cycle, the most efficient use of the sodium atoms will be with repumping and pulse durations between 10–30 μs. The time between these long pulses (50–150 μs) would be sufficient for pulse-to-pulse effects to be negligible (Rampy et al. 2013). However, if partitioning of the laser light to include the D$_2$b line is not possible due to engineering constraints, somewhat shorter pulses will be better suited to take advantage of the initial return enhancement.

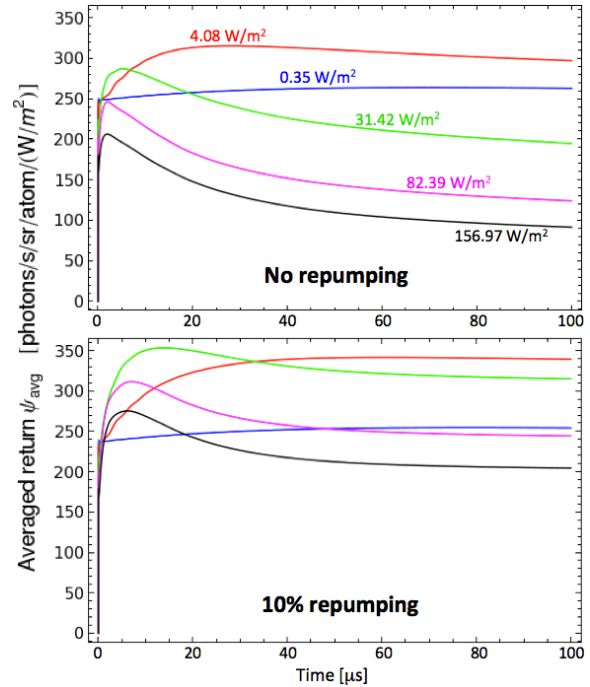

Fig. 2. Time evolution of specific return (in the mesosphere) for different levels of irradiance present for 10 W average power and 20% duty cycle over 100 μs. When no light is resonant with D$_2$b (top plot), the return peaks and declines quickly. When 10% is tuned to D$_2$b (bottom plot), the return peaks later, is higher, and decreases less rapidly. These graphs show that certain values of irradiance excite the sodium atoms more efficiently.

It was hypothesized that the peak and decline are in part due to the effects of Larmor precession, since this acts on timescales of a few microseconds. To test this, these simulations were repeated with zero magnetic field. The results showed a lessening in the severity of the decline, and the expected increase in return, but the peak remained prominent. Therefore, other factors are of greater importance, such as the rate of collisions between the sodium atoms and other constituents of the mesosphere. Indeed, altering the rate of collisions changes the shape of the curves in Figure 2, and the predicted return flux, as is discussed below.

Another interesting observation from Figure 2 is that certain irradiance levels are more efficient at optical pumping than others. In the bottom pane of Figure 2 (for 10% repumping) the green line corresponding to 31.4 W/m$^2$ achieves the highest specific return. The possibility of taking advantage of this circumstance by controlling the irradiance profile using uplink correction will be discussed further in Section 4.

Beyond the consideration of how to most effectively excite the sparse sodium atoms during a single pulse, is the question of what pulse formats allow for the benefits of optical pumping to be maintained between consecutive pulses. Stated differently, what is the length of time necessary for atomic polarizations to return to thermal equilibrium? Based on results of various simulations performed to test this, the time scale appears to be on the order of ~50 μs, near the center of the sodium layer. Hence, a format with less than this amount of dark time between pulses should be modeled with multiple cycles of the laser, while pulses with larger spacing may be considered independent, i.e., modeling a single pulse will give an estimate within 5 to 10% what is obtained for multiple pulses.

Figure 3 shows a 10-pulse train for pulses of 500 ns, with 10% repumping and a 20% duty cycle. There is collisional relaxation and Larmor precession between pulses, but the ground-state polarization (and possibly also the non-thermal velocity distribution) created by the light during the pulse has not completely relaxed by the time the next pulse comes. The pumping between pulses is evident. Notably, the different irradiance levels (which are the same as in Figure 2) evolve differently as a function of time. The extremal irradiance levels (0.35 and 156.97 W/m$^2$) remain fairly constant from pulse to pulse, while the in between irradiances (especially the 31.42 and 82.39 W/m$^2$ curves, in green and magenta, respectively) show the highest return after just a few pulses. These irradiances are most efficient at exciting the sodium atoms.

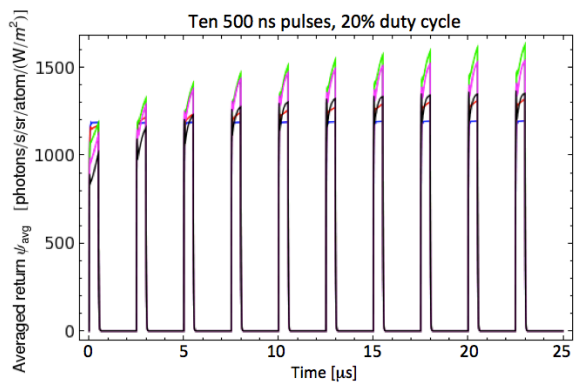

Fig. 3. This intermediate pulse format shows effects of optical pumping between pulses, indicating the sodium atoms maintain their polarization state during the 2 μs dark time. The irradiance levels and colors are the same as in Figure 2. The green (31.42 W/m$^2$) and magenta (82.39 W/m$^2$) levels have the highest specific return after a few pulse cycles, indicating they are more efficient than the other levels.

To discover the optimal pulse duration, simulations were carried out for pulses ranging from 200 ns to 60 μs, with 10% repumping, both 10% and 20% duty cycles, for zenith pointing at Mt. Hamilton. In all cases, the 20% duty cycle gave higher return, and the maximum was for the 30 μs pulses. The predicted return flux at the top of the telescope (i.e. before interacting with any optics or detectors) for this format is 12 × 10$^6$ photons/s/m$^2$, assuming a sodium column density of 4 × 10$^{13}$ atoms/m$^2$ and 0.5 m FWHM mesospheric spot size. This value is based on simulation of 3 cycles of the laser (450 μs), although it is the same value obtained for only one cycle. At a pulse length of 60 μs, and the same duty cycle, the time-averaged return flux declined to 11.2 × 10$^6$ photons/s/m$^2$. For the 200 ns case (the LLNL fiber laser is currently operating with this pulse length, but at 10% duty cycle), the predicted return flux is 11.3 × 10$^6$ photons/s/m$^2$ based on a simulation of 50 cycles (50 μs). This information is displayed in Table 1. For comparison, a 10 Watt CW laser with otherwise identical parameters (repump fraction, location and pointing) is expected to produce 13 × 10$^6$ return photons/s/m$^2$.

Table 1. Predicted return flux at the telescope for 200 ns, 30 μs, and 60 μs pulses, with 10% and 20% duty cycles, from a zenith pointing 10 W laser at Mt. Hamilton assuming a sodium column density of 4 × 10$^{13}$ atoms/m$^2$ and 0.5 m FWHM mesospheric spot.

| Pulse length (μs) | Return 10% duty cycle (×10$^6$ photons/s/m$^2$) | Return 20% duty cycle (×10$^6$ photons/s/m$^2$) |
|---|---|---|
| 0.2 | 9.5 | 11.3 |
| 30 | 10.5 | 12.0 |
| 60 | 10.2 | 11.2 |

The default time between collisions of sodium atoms and other constituents of the upper atmosphere (mainly $O_2$ and $N_2$) in LGSBloch is 35 μs for velocity changing and 245 μs for spin randomizing. However, these values are estimates, due to the fact that the relevant cross sections are not well known. To test the effect of changes in this variable on the predicted return flux, the case of 30 μs pulses and 20% duty cycle was examined for collision times 10 times shorter and 10 times longer, in the absence of a magnetic field. The results showed an ~12% decrease in return flux at the telescope when the rate of collisions were an order of magnitude smaller than the default, and only an ~1% increase when collisions were an order of magnitude less frequent. This determines that LGS brightness is not highly disturbed by fluctuations in this parameter.

The above investigation established the 30 μs pulses with 20% duty cycle as part of the goal format to implement on the LLNL fiber laser, since it has the highest predicted return and will also facilitate Raleigh blanking. The 30 μs pulse length corresponds to a distance of 9 km in air, so the pulse will essentially span the entire sodium layer.

The return flux predictions given thus far have been for narrow (<10 MHz) line width lasers. The spectrum of the LLNL fiber laser currently contains multiple lines (modes) spaced at 200 MHz intervals. This multimode format was implemented to mitigate stimulated Brillouin effects (SBS) in the fiber amplifiers. From the standpoint of effective coupling to the Na atoms, in the absence of saturation such a spectrum is generally less desirable than having the light confined within two lines that are tuned precisely to the $D_2$a and $D_2$b transitions. Also, the problem of accurately modeling a system with many discrete frequency components using Bloch equations becomes very complex. Hence, the numbers presented thus far should be considered optimistic upper limits on what will actually be achievable with this laser in real on-sky conditions.

There is some flexibility in the fiber laser system as to the number of modes and how far apart they are spaced, and the option exists to convert to a broadband format instead. Because the freely available version of LGSBloch does not have the capability to do multimode time domain calculations, a consulting contract was established with one of the co-authors (SR) through his consulting firm Rochester Scientific to investigate the effects of such changes on the return flux.

Rochester Scientific developed a proprietary extension of the LGSBloch simulation package, where solutions to the density matrix evolution equations are generated with a code written in C that uses an open-source ODE solver (Rochester et al. 2012). This method addresses the stiffness of the system using implicit backward differentiation formulas (Hindmarsh et al. 2005), and significantly decreases the time required for convergence to a solution. Techniques to account for the atoms being subjected to a field with components oscillating at different frequencies have also been included.

Hillman et al. (2006) hypothesized that a mode separation of 150 MHz would give maximal return flux, based on considerations of the sodium hyperfine structure. To investigate this, Rochester Scientific produced a simulation for a three-mode system, with the central frequency tuned to the center of the D₂a line and sidebands with separations ranging from 0 to 500 MHz. The parameters used were for a 10 W laser at Mt. Hamilton with 30 μs pulses, 20% duty cycle, and no repumping. The orange curves in Figure 4 show how the specific return in the mesosphere varies with mode separation, when propagating parallel (top) and perpendicular (bottom) to the geomagnetic field. The blue curves show the changes in return for a broadband laser spectrum, with line width ranging between 0 and 500 MHz.

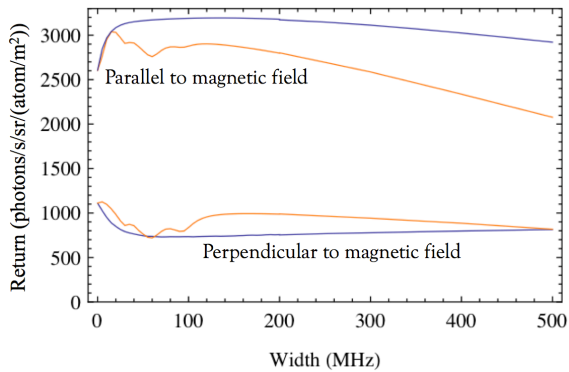

Fig. 4. The predicted return in the mesosphere for a 10 W laser at Mt. Hamilton with 30 μs pulses, 20% duty cycle, and no repumping, varies depending on the spectral format. Return is shown for a multimode LGS as a function of mode separation (orange) and for a broadband LGS as a function of bandwidth (blue), with the laser directed parallel (top lines) and perpendicular (bottom) to Earth's magnetic field. Different formats may be preferable at different locations depending on the orientation to the magnetic field during the majority of observations.

The structure observed in the multimode case can be understood in terms of the energy-level structure of sodium. The local minimum at ~60 MHz is a consequence of the 59.8 MHz separation between the $F' = 2$ and $F' = 3$ excited states. Here, while one of the laser modes (say, the central frequency) is exciting the $F = 2$ to $F' = 3$ transition, an adjacent mode (say, one of the sidebands) is tuned to the $F = 2$ to $F' = 2$ transition for the same velocity group. Because atoms in the $F' = 2$ excited state can decay to the $F = 1$ ground state where they are no longer resonant with the light field, excitations of the $F' = 2$ state are less beneficial, and the competition with optical pumping on the $F = 2$ to $F' = 3$ transition leads to reduced return flux The dips in return near ~30 and ~90 MHz can be understood similarly. When the sidebands are 30 MHz from the central frequency, atoms in a velocity group resonant with the $F = 2$ to $F' = 3$ transition for one sideband will also be resonant with the opposite sideband (spaced 60 MHz away) exciting the $F = 2$ to $F' = 2$ transition. When the modes are separated by 95.3 MHz, a velocity group resonant with a laser mode exciting the $F = 2$ to $F' = 3$ transition will also be resonant with an adjacent mode exciting the $F = 2$ to $F' = 1$ transition. Atoms in the $F' = 1$ excited state can also decay to the $F = 1$ ground state and be lost, leading to reduced return flux due to the competition, as above. The decline in return at larger separations is attributed to the presence of fewer atoms available to interact with the light field near the edges of the Doppler broadened spectrum.

Figure 4 confirms that the ideal line separation in a multimode LGS is either small (less than ~20 MHz) or between about 120 to 150 MHz, depending on the orientation to Earth's magnetic field. This holds implications for what will be the most effective use of laser power depending on the location of an observatory, and what the highest priority science cases are. Near the magnetic equator, a laser may spend the majority of time propagating near perpendicular to the geomagnetic field. At more magnetically polar latitudes, zenith propagation can be much closer to parallel to the field lines. Figure 4 suggests that a broadband LGS (with FWHM between about 50 and 150 MHz) should be the brightest for polar observatories, and an equatorial location may benefit most from a multimode spectrum (with mode separations of ~150 MHz).

Taking into account the lessons learned from the inquiries discussed in this section, the goal format for the LLNL fiber laser is presented in Table 2.

Table 2. Comparison of parameters between current Mt. Hamilton LGS, and lab tested and goal formats of the new laser

|  | Current dye laser | New fiber laser | |
| --- | --- | --- | --- |
|  |  | Lab tested format | Goal format |
| Output power | 9 W | 10 W | 10 W |
| Polarization | Linear | Circular | Circular |
| Spectral format | ~2 GHz FWHM bandwidth | 9 lines with 200 MHz spacing | Fewer lines and/or smaller spacing |
| Pulse duration | 150 ns | 200 ns | 30 μs |
| Duty cycle | 0.16 % | 10 % | 20 % |
| Fraction of light on D₂b | None | None | 10% |

## 4. POTENTIAL BENEFITS OF UPLINK CORRECTION

Numerical simulations enable investigation of the possible benefits of uplink control, a technique in which the laser light is reflected off a deformable mirror prior to projection into the atmosphere (Norton et al. 2014). This allows manipulation of the wavefront in order to compensate for imperfections of the outgoing beam shape and projection optics, and aberrations induced by atmospheric turbulence. The potential benefits explored include obtaining a smaller spot to decrease WFS measurement error, and potentially regulating the pattern of irradiance in the mesosphere to maximize efficiency of optical pumping.

In consulting with Rochester Scientific, the issue of quantifying the benefits of uplink control was explored. We investigated manipulating the irradiance profile as well as the spot size, in order to maximize the area in the mesosphere irradiated at levels that lead to the most efficient fluorescence (as discussed in Section 3). This involved modeling a beam profile with the shape of a top hat, and comparing it to the Gaussian profiles used for all modeling results thus far.

In Figure 5, two curves compare the predicted return flux at the top of the telescope of both Gaussian and top hat shaped mesospheric illumination profiles. These were generated for a laser that produces 10 W average power, located at Mt. Hamilton, operating at 20% duty cycle with 30 μs pulses, 3 modes spaced by 150 MHz, 10% of the power in a sideband tuned to D₂b, and another 10% lost in the sideband on the far side of the spectrum (this is necessary as a consequence of current laser engineering frequency modulation techniques). The assumed sodium column density is 4 × 10¹³ atoms/m².

We can conclude that there is only a small enhancement in return flux over the Gaussian beam when the irradiance profile is made to approach a constant value. However, full interpretation requires understanding how the shape of the beam in the mesosphere gets translated to a spot size on the WFS detector. For most LGS AO systems currently in operation, there are advantages to using uplink correction to reduce the size of the laser spot.

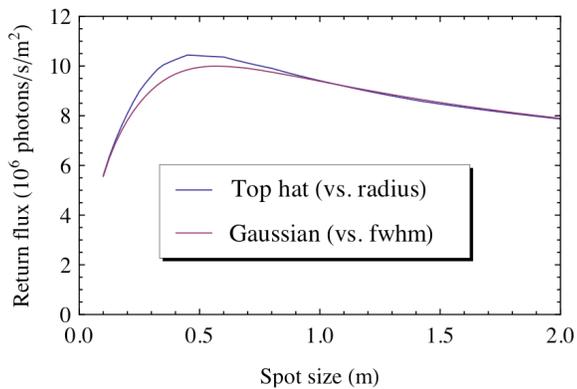

Fig. 5. The predicted return flux at the top of the telescope for Gaussian and top hat shaped mesospheric irradiance profiles shows slight improvement with the top hat profile. This is for the case of 10 W average power, at Mt. Hamilton, 20% duty cycle, 30 μs pulses, 3 modes spaced by 150 MHz, 10% of the power in a sideband at D$_2$b, and another 10% lost on the far side of the spectrum. The largest benefits of uplink correction will come from the reduction in WFS measurement error associated with a smaller spot size.

The current estimate of the FWHM spot size of the LGS on the Shane telescope is ~2 arcseconds, which means it appears to be ~1 m diameter at the distance of the mesosphere (90 km). Assuming that there are approximately equal amounts of spot broadening from atmospheric turbulence as the light travels back to the telescope as for its trip upward, then the actual physical size of the laser beam in the mesosphere is ~0.6 m. If uplink control could be used to reduce this by a factor of two to a nearly diffraction limited spot, even with the decreased return as indicated in Figure 5, the WFS measurement error would be reduced by a factor of two. Such benefits may be worth the additional complexity of an uplink system. Recall that LGS AO wavefront measurement performance is tied to the guide star radiance (energy per unit area at the telescope), not just the total brightness, so some minor sacrifice of return flux is tolerable if the apparent spot size is reduced.

The best estimate of the LGS brightness expected from the LLNL fiber laser can be deduced from Figure 5. Assuming a roughly Gaussian spot that has ~0.6 m FWHM, approximately 10 × 10$^6$ photons/s/m$^2$ should reach the top of the telescope. This is somewhat less than the 12 × 10$^6$ photons/s/m$^2$ predicted in section 3, but the difference can be understood as a consequence of the more realistic spectral format used here. Given that the average measured return flux from the current dye laser on the Shane telescope between 2006 and 2010 was 1.16 × 10$^6$ photons/s/m$^2$, it appears reasonable to expect the fiber laser to produce 5–10 times the return seen from the current LGS.

## 5. COMPARISON OF SIMULATION PREDICTIONS TO LGS RETURN MEASUREMENTS

Because the return flux at the telescope scales linearly with the mesospheric sodium column density, $C_{Na}$, this parameter is a useful gauge in determining how accurately a numerical model reproduces real systems. Results of LIDAR measurements indicate the expected median annual density is 3–4 × 10$^{13}$ atoms/m$^2$, although sporadic layers may have concentrations up to 10 times greater (Xiong et al. 2003, Moussaoui et al. 2010). It has long been known that the sodium layer has variations that are seasonal, rather than annual, with a winter abundance maximum in both the northern and southern hemispheres (Plane et al. 1999). Perhaps due to the varying intensity of seasonal temperature changes, the strength of variability is a function of latitude, with more extreme changes at larger latitudes (Gumbel et al. 2007).

The LGSBloch model is reported to predict return fluxes in good agreement with measurements for the case of the single frequency 50 W Fasortronics laser at SOR (Holzlohner et al. 2010b). This requires assuming slightly higher than average, but still reasonable, values for the $C_{Na}$, which are fairly consistent with what has been reported for that site (Drummond et al. 2007). Also, good agreement is seen between model predictions and measurements for the CW laser used in the ESO transportable LGS unit (Bonaccini Calia et al. 2012).

The parameters of the current LGS on the Shane telescope at Mt. Hamilton are given in Table 2. Because of the long dark time between pulses (~91 μs), and long computation times required by LGSBloch, initial modeling endeavors only included one cycle of the laser. They incorporated a Gaussian spot size of 0.6 m FWHM in the mesosphere, a Gaussian pulse shape with 150 ns FWHM, and a Gaussian shaped spectrum with 2.2 GHz FWHM bandwidth. The values of these last two parameters were determined from measurements taken with Mt. Hamilton instruments, although the shape of the laser spectrum is closer to a top hat.

In Figure 6, return flux measurement data from the AO system WFS of this laser, taken between 2006 and 2010, are plotted as a function of month. They show strong evidence for the seasonal variation of the sodium layer. The value of $C_{Na}$ found by comparing the measured data with the prediction from the model was used as a gauge to how well the measurements were reproduced with LGSBloch. The right-hand axis exhibits the implied amount of $C_{Na}$ based on the present stage of modeling. For the seasonal average corresponding to a return flux of 1.16 × 10$^6$ photons/s/m$^2$ at the top of the telescope, the average $C_{Na}$ is found to be only 1.5 × 10$^{13}$ atoms/m$^2$. This small value is interpreted as an indication that the simulations are over-predicting the measurements for this laser by at least a factor of two. Thus, the conclusion is that some physical, environmental, or observational parameters are not being accounted for adequately.

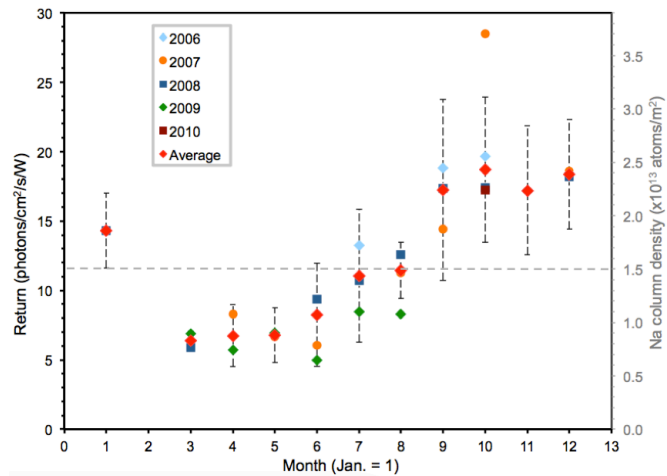

Fig. 6. Data from the dye laser on the Shane telescope show expected seasonal variability. The right axis shows implied $C_{Na}$ based on current modeling with LGSBloch. The horizontal dashed line indicates the predicted average column density for the measured average return of 11.6 photons/cm$^2$/s/W. These values are considerably lower than expected.

The return flux measurements in Figure 6 are based on data collected with the Shack-Hartman WFS, for zenith pointing observations. The intensity in the fully illuminated subapatures was averaged, and corrected for detector effects and 40% system throughput. The camera gain is reported to be ∼1 electron/count. In 2004, a portable power meter with 633 nm wavelength light was used to measure throughput of the AO system from the primary focus of the telescope to the WFS, which was found to be 55%. In January 2012, observations of several Hayes spectrophotometric standard stars were taken by the observatory staff to re-asses this value. A sodium filter with 0.87 nm FWHM bandpass was placed in the beam path, and resulting photometry measurements were corrected for atmospheric transmission and compared with online spectra[†]. This analysis yielded an average throughput of 40%, which is consistent with the 2004 measurement when considering the amount of light lost to the primary and secondary mirror surfaces and optics internal to the WFS.

After ensuring the measurement data were being interpreted correctly, some of the less precisely determinable simulation inputs were varied to see how sensitive the predictions were to these assumptions. Making adjustments to parameters such as spot size, bandwidth, and atmospheric transmission, it is possible to somewhat ameliorate the disagreement, but only marginally. For example, the estimate of return for the Shane laser can be reduced 20% by reducing the modeled spot size from 0.7 m to 0.6 m FWHM, and changing the bandwidth from 1.5 GHz (the original target width for the spectrum) to 2.2 GHz FWHM (closer to the measured value). However, the same reduction also occurred when increasing the spot size to 1.0 m FWHM, setting the bandwidth to 2.2 GHz, and decreasing the atmospheric transmission to 85% (the default value is 89%). This is strong evidence for the need to better constrain such variables.

It is worth noting a check of the relative accuracy of LGSBloch was performed by computing the predicted return when the spectrum broadening phase modulators were switched off. In both simulations and experiment this reduced the LGS brightness by a factor of ∼3 (presumably due to increased saturation of the sodium layer).

Additional efforts to improve the agreement between theory and measurement were explored when consulting with Rochester Scientific. This entailed reduction of the launched power to 9 W, simulation of multiple cycles of the laser, and inclusion of a more realistic laser spectrum and mesospheric irradiance profile. However, these changes were not able to ameliorate the discrepancy (Rampy 2013).

To help identify whether the source of the discrepancy was related to the measurement data or the model, simulations were also performed for the two LGS at the Keck observatory. The Keck II system began science operations in 2004 (Wizinowich et al. 2006), and has been responsible for ∼70% of the refereed science papers published worldwide through 2014 based on data collected with LGS AO systems. The success of this system led to the desire for similar capabilities on the Keck I telescope. This has been recently realized, with the Keck I LGS AO system having begun full science operations in the latter part of 2012 (Chin et al. 2012).

The laser on Keck II is the same technology as the dye laser at the Lick observatory and has essentially the same format, except that it operates with a repetition rate of 26 kHz (the Shane laser is at 11 kHz). A Gaussian spot size of 0.7 m FWHM in the mesosphere was assumed, and parameters based on location, such as observatory altitude, magnetic field direction and strength, were included in the model. For the ∼12 W expected to be projected from the launch telescope, the simulations predict $4.2 \times 10^6$ photons/s/m², for a sodium column density of $4 \times 10^{13}$ atoms/m². However, measurements reported in the literature indicate between ∼0.7 and $1.75 \times 10^6$ photons/s/m² are being received at the telescope (Le Mignant et al. 2006). This again constitutes a factor of 2–3 discrepancy between data and simulation.

The Keck I laser was developed by Lockheed Martin Coherent Technology (LMCT) and is a solid-state mode-locked CW laser, which uses sum-frequency generation (SFG) of beams at 1064 nm and 1319 nm to achieve the 589 nm sodium wavelength (Sawruk et al. 2010). The efficiency of the SFG process increases nonlinearly with input light intensity, so to maximize efficiency the laser is formatted as a train of short pulses to achieve high peak intensity. Each pulse is 300 picoseconds FWHM and the repetition rate is 82 MHz. Though the average launched power is higher for this laser (∼ 18 W), the return from the sodium layer is on average nearly equal to, or only slightly higher, than what is seen from the Keck II LGS (Chin et al. 2012). LGSBloch simulations give the over-prediction that $5.2 \times 10^6$ photons/s/m² should be reaching the top of the telescope. All evidence suggests that this format is even less effective at coupling to the sodium atoms than those used by the dye lasers.

## 6. CAN PULSED LASERS PRODUCE BRIGHTER LGS THAN CW?

A recent suggestion has raised the possibility that pulsed systems could produce brighter beacons than are achievable with CW lasers. The scheme entails pulsing the laser at the Larmor frequency in order to avoid the detrimental effects of atomic states changing their magnetic quantum number as they precess (Hillman et al. 2012). This is similar to an idea in the field of magnetometry, to use LGS systems with modulated lasers to measure the strength of Earth's magnetic field in the mesosphere (Budker et al. 2007).

Figure 7 presents a preliminary assessment of this technique, for a 10 W average power LGS at Mt. Hamilton (solid lines) and 20 W average power at Mauna Kea (dashed lines). At an altitude of 90 km above Mt. Hamilton, the magnetic field strength and Larmor precession period are $B = 0.47\,G$, $\tau_L \approx 3.1$ μs (326 kHz), and above Mauna Kea $B = 0.33\,G$ and $\tau_L \approx 4.4$ μs (231 kHz). Pulsed simulations included these values as the repetition rates, with 10% duty cycle, and 10% repumping, for the laser beam pointing parallel and perpendicular to the magnetic field, and towards zenith, for each site. The return flux is compared to CW formats with the same average power and otherwise identical parameters, and also to the goal format in the case of Mt. Hamilton. All simulations included five irradiance levels, integrated over a 0.5 m FWHM Gaussian beam profile in the mesosphere, and propagated back through the atmosphere to the top of the telescope.

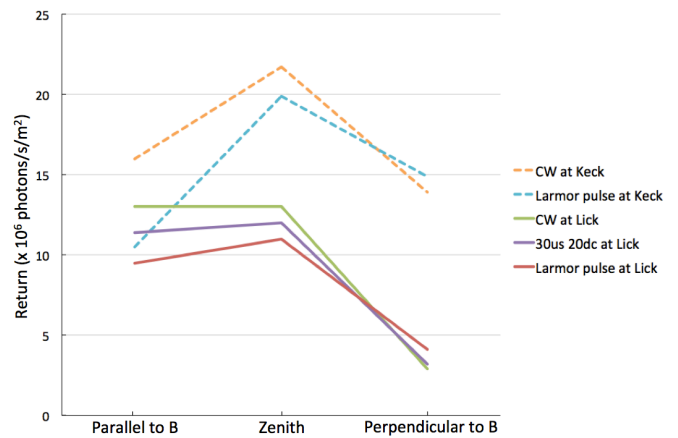

Fig. 7. Comparisons of return flux for 10 W average power at Mt. Hamilton (solid) and 20 W average power at Mauna Kea (dashed) for the Larmor pulse format, a CW format, and at Mt. Hamilton, the goal

---

[†] www.eso.org/sci/observing/tools/standards/spectra/stanlis.html

format for the LLNL fiber laser. Simulations were carried out for a laser pointing parallel and perpendicular to the magnetic field, and to zenith.

The largest enhancement is seen for the case of propagation perpendicular to the magnetic field at Mt. Hamilton, where the Larmor frequency pulses produce ~40% higher return than a CW laser with the same average power. At Mauna Kea there is only a 7% increase over CW, likely because the strength of the geomagnetic field is less at this location. These results suggest implementing Larmor frequency pulses at a fairly equatorial site such as Mauna Kea is unlikely to produce significant benefits. However, at the location of Mt. Hamilton, and depending on the science cases of greatest interest, such a technique may warrant further inquiry.

To ensure the repetition frequency used was indeed at the peak of the resonance, several slightly different repetition rates were simulated. This investigation showed that the peak is narrow (only a few kilohertz wide) for a given height in the sodium layer, so experimental investigation of this phenomenon may require the ability to precisely tune the repetition rate of the laser. Future investigation of this method will focus on determining how this enhancement can be most effectively used to increase LGS brightness.

## 7. DISCUSSION

Sodium laser guide stars (LGS) have emerged as an important tool for advancing ground-based astronomy. The significant cost of these systems necessitates maximizing efficiency, which is an effort that is still in progress. The LGSBloch simulation package (available at http://rochesterscientific.com/ADM/) has permitted new insights into how laser light interacts with mesospheric sodium, including the evolution of the energy level populations, and other physical effects such as Larmor precession, atomic recoil, and collisions with other constituents of the upper atmosphere. Guidelines for current and future system designers have been identified through use of this tool.

Predicted return fluxes from the LGSBloch simulation package have been shown to be reasonably accurate for the case of CW LGS, but to overestimate the return of short pulse systems by a factor of 2–3. The two cases of CW LGS are the 50 Watt Fasortronics laser at the Starfire Optical Range and ESO transportable LGS unit. For short pulse systems, we have modeled the lasers at the Lick and Keck Observatories, and a similar discrepancy was reported for comparison to the LGS system at the Gemini South Observatory (Holzlohner et al. 2012).

The consistency of the discrepancy between four short pulse laser systems with independent assessments raises suspicion that the issue lies in our current understanding of this format regime rather than a measurement calibration error. Because the formats investigated for the new fiber laser at Mt. Hamilton have intermediate pulse lengths (microseconds, instead of nano- or pico-seconds), it is expected the actual return should be less than a factor of 2 reduced from the LGSBloch predictions, assuming its format matches the goal format presented here.

Pulsed laser systems require substantially more complex simulations, and it is possible the discrepancy could be reduced with improved modeling of velocity-changing collisions or other time domain phenomena. It has also been suggested that the problem could be due to some unknown fraction of the light getting lost, or diverted to non-interacting frequencies, between the diagnostic measurements and laser launch. The likeliness of this has been ascertained from plots of the expected return flux as a function of launched laser power. These indicate that ~70% of the power would have to be unaccounted for (assuming the other parameters used are reasonably accurate), a situation that does not seem probable. Other parameters it would be beneficial to double check experimentally include:
- Wavelength stabilization
- Launched power and polarization state
- Beam quality and corresponding irradiance histogram in the mesosphere
- Atmospheric transmission at the time of measurements

An important step in improving our understanding of sodium-light interactions will be accumulating additional experimental information. Dedicated tests to evaluate the effect of changing the power and spectrum of the laser would help illuminate the specific areas of our understanding that require improvement.

## References

1. Happer, W., MacDonald, G. J., Max, C. E., and Dyson, F. J., "Atmospheric-turbulence compensation by resonant optical backscattering from the sodium layer in the upper atmosphere," J. Opt. Soc. Am. A 11, 263–276 (1994).
2. Holzlöhner, R., Rochester, S. M., Calia, D. B., Budker, D., Higbie, J. M., and Hackenberg, W., "Optimization of cw sodium laser guide star efficiency," Astron. Astrophys. 510, A20+ (2010a).
3. Kupke, R., Gavel, D., Roskosi, C., Cabak, G., Cowley, D., Dillon, D., Gates, E. L., McGurk, R., Norton, A., Peck, M., Ratliff, C., and Reinig, M., "ShaneAO: an enhanced adaptive optics and IR imaging system for the Lick Observatory 3-meter telescope," Proc. SPIE 8447, Adaptive Optics Systems III, 84473G (September 13, 2012).
4. Hillman, P. D., Drummond, J. D., Denman, C. a., and Fugate, R. Q. "Simple Model, Including Recoil, for the Brightness of Sodium Guide Stars Created from CW Single Frequency Fasors and Comparison to Measurements" Proc. SPIE 7015, Adaptive Optics Systems, 70150L (2008).
5. Jeys, T.H., Heinrichs, R. M., Wall, K. F., Korn, J., Hotaling, T. C., and Kibblewhite, E., "Observation of optical pumping of mesospheric sodium," Optics Letters, Vol. 17, No 16 (1992).
6. Milonni, P. W., Fugate, R. Q., and Telle, J. M., "Analysis of measured photon returns from sodium beacons," J. Opt. Soc. Am. A 15, 217–233 (1998).
7. Gumbel, J., Fan, Z. Y., Waldemarsson, T., Stegman, J., Witt, G., Llewellyn, E. J., She, C. –Y., and Plane, J. M. C., "Retrieval of global mesospheric sodium densities from the Odin satellite," Geophysical Research Letters, Vol. 34, L04813 (2007).
8. Holzlöhner, R., Rochester, S. M., Calia, D. B., Budker, D., Pfrommer, T., and Higbie, J. M., "Simulations of pulsed sodium laser guide stars: an overview," Proc. SPIE 8774, Adaptive Optics Systems III, 84470H (2012).
9. Corney, A., in "Atomic and Laser Spectroscopy," Clarendon, Oxford, (2006).
10. Wang, L., Otarola, A., Ellerbroek, B., "Impact of sodium laser guide star fratricide on multi-conjugate adaptive optics systems," JOSA A., Vol. 27, No. 11, pp. A19-A28 (2010).
11. Bonaccini-Calia, D., Hackenberg, W., Holzlöhner, R., Lewis, S., and Pfrommer, T., "The Four-Laser Guide Star Facility: Design considerations and system implementation," Adv. Opt. Techn., 3,3, pp. 345-361 (2014).
12. Rampy, R., "Advancing adaptive optics technology: Laboratory turbulence simulation and optimization of laser guide stars," Dissertation, University of California Santa Cruz (2013).
13. Rochester, S. M., Otarola, A., Boyer, C., Budker, D., Ellerbroek, B., Holzlöhner, R., and Wang, L., "Modeling of pulsed-laser guide stars for the Thirty Meter Telescope project," JOSA B, Vol. 29, Issue 8, pp. 2176-2188 (2012).
14. Hindmarsh, A. C., Brown, P. N., Grant, K. E., Lee, S. L., Serban, R., Shumaker, D. E., and Woodward, C. S., "SUNDIALS: Suite of nonlinear and differential/algebraic equation solvers," ACM Transactions on Mathematical Software 31, 363–396 (2005).
15. Hillman, P., "Sodium guidestar return from broad CW sources," CfAO Laser Workshop (2006).
16. Norton, A. P., Gavel, D., Helmbrecht, M., Kempf, C., Gates, E., Chloros, K., Redel, D., Kupke, R., and Dillon, D., "Laser guidestar uplink correction using a MEMS deformable mirror: on-sky test results and implications for future AO systems," Proc. SPIE 9148, Adaptive Optics Systems IV, 91481C (2014).

17. Xiong, H., Gardner, C. S., and Liu, A. Z., "Seasonal and nocturnal variations of the mesospheric sodium layer at Starfire Optical Range, New Mexico," Chin. J. of Geophys. 46, pp. 432–437 (2003).
18. Moussaoui, N., Clemesha, B.R., Holzlöhner, R., Simonich, D. M., Calia, D. B., Hackenberg, W., and Batista, P. P., "Statistics of the sodium layer parameters and its impact on AO sodium LGS characteristics," Astron. Astrophys. 511, A31 (2010).
19. Plane, J. M. C., Gardner, C. S., Yu, J. R., She, C. Y., Garcia, R. R., and Pumphrey, H. C., "Mesospheric Na layer at 40°N: Modeling and observations," J. Geophys. Res., 104, 3773 (1999).
20. Holzlöhner, R., Rochester, S. M., Pfrommer, T., Calia, D. B., Budker, D., Higbie, J. M., and Hackenberg, W., "Laser guide star return flux simulations based on observed sodium density profiles," Proc. SPIE 7736, Adaptive Optics Systems II, 77360V (2010b).
21. Drummond, J., Novotny, S., Denman, C., Hillman, P., and Telle, J., "The Sodium LGS Brightness Model over the SOR," AMOS 2007 Conference, pp. E67 (2007).
22. Bonaccini Calia, D., Guidolin, I., Friedenauer, A., Hager, M., Karpov, V., Pfrommer, T., Holzlohner, R., Lewis, S., Hackenberg, W., Lombardi, G., Centrone, M., and Pedichini, F., "The ESO transportable LGS Unit for measurements of the LGS photon return and other experiments", Proc. SPIE 8450, Modern Technologies in Space- and Ground-based Telescopes and Instrumentation II, 84501R (2012).
23. Wizinowich, P. L., Le Mignant, D., Bouchez, A. H., Campbell, R. D., Chin, J. C. Y., Contos, A. R., van Dam, M. A., Hartman, S. K., Johansson, E. M., Lafon, R. E., Lewis, H., Stomski, P. J., and Summers, D. M., "The W. M. Keck Observatory Laser Guide Star Adaptive Optics System: Overview," PASP, 118, 297-309 (2006).
24. Chin, J. C. Y., Wizinowich, P., Campbell, R., Chock, L., Cooper, A., James, E., Lyke, J., Mastromarino, J., Martin, O., Medeiros, D., Morrison, D., Neyman, C., Panteleev, S., Stalcup, T., Tucker, P., Wetherell, E., and van Dam, M., "Keck I Laser Guide Star Adaptive Optics System," Proc. SPIE 8447, Adaptive Optics Systems III, 84474F (2012).
25. Le Mignant, D., Campbell, R., Boucheza, A., Chin, J., Chock, E., Conrad, A., van Dam, M., Doyle, S., Goodrich, R., Johansson, E., Lafon, R., Lyke, J., Melcher, C., Mouser, R., Summers, D., Wilburn, C., and Wizinowich, P., "LGS AO operations at the W. M. Keck Observatory," Proc. SPIE 6270, Observatory Operations: Strategies, Processes, and Systems, 62700C (2006).
26. Sawruk, N. W., Lee, I., Jalali, M. P., Prezkuta, Z., Groff, K. W., Roush, J., Rogers, N., Tiemann, B., Hannon, S. M., Alford, W. J., d'Orgeville, C., Fesquet, V., Oram, R., Adkins, S. M., and Grace, K., "System overview of 30 W and 60 W 589 nm guidestar laser systems," SPIE Proc. 7736, Adaptive Optics Systems II, 77361Y (2010).
27. Hillman, P., "Paths towards a brighter guidestar," CfAO Laser Workshop (2012).
28. Budker, D., and Romalis, M. V., "Optical Magnetometry," Nature Physics 3, 227 - 234 (2007).